\documentclass[12pt,twoside,a4paper]{article}
\usepackage{amsmath,amssymb,latexsym,theorem,natbib,epsfig,color,subfigure,footmisc,bbm}
\usepackage{multirow,graphicx,array,rotating,url,multicol}  
\usepackage{epstopdf,afterpage,wasysym,hyperref}
\usepackage{verbatim}
\usepackage[normalem]{ulem}
\def\logs{{\hbox{\rm LogS}}}
\def\rel{\mathop{\hbox{\rm Rel}}}
\def\shp{\mathop{\hbox{\rm Shp}}}

\newcommand{\degree}{\ensuremath{^\circ}}

\title{Ensemble size dependence of the logarithmic score for forecasts issued as multivariate normal distributions}

\author{{Martin Leutbecher}$^{1}$ and {S\'andor Baran}$^{2,*}$ \vspace*{0.5cm}\\
{\small $^1$European Centre for Medium-Range Weather Forecasts, Reading, United Kingdom}\\
{\small $^2$Faculty of Informatics, University of Debrecen, Hungary}}

\date{}

\begin{document}
\maketitle

\footnotetext[1]{Corresponding author: \url{baran.sandor@inf.unideb.hu}}

\begin{abstract}
Multivariate probabilistic verification is concerned with the evaluation of joint probability distributions of vector quantities such as a weather variable at multiple locations or a wind vector for instance. The logarithmic score is a proper score that is useful in this context. In order to apply this score to ensemble forecasts, a choice for the density is required. Here, we are interested in the specific case when the density is multivariate normal with mean and covariance given by the ensemble mean and ensemble covariance, respectively. Under the assumptions of multivariate normality and exchangeability of the ensemble members, a relationship is derived which describes how the logarithmic score depends on ensemble size. It permits to estimate the score in the limit of infinite ensemble size from a small ensemble and thus produces a fair logarithmic score for multivariate ensemble forecasts under the assumption of normality. This generalises a study from 2018 which derived  the ensemble size adjustment of the logarithmic score in the univariate case.  

An application to medium-range forecasts examines the usefulness of the ensemble size adjustments when multiva\-riate normality is only an approximation. Predictions of vectors consisting of several different combinations of upper air variables are considered. Logarithmic scores are calculated for these vectors using ECMWF's daily extended-ran\-ge forecasts which consist of a  100-member ensemble. The probabilistic forecasts of these vectors are verified aga\-inst operational ECMWF analyses in the Northern mid-lati\-tudes in autumn 2023. Scores are computed for ensemble sizes from 8 to 100. The fair logarithmic scores of ensembles with different cardinalities are very close, in contrast to the unadjusted scores which decrease considerably with en\-sem\-ble size. This provides evidence for the practical usefulness of the derived relationships.

\bigskip
\noindent {\em Keywords:\/} {ensemble forecast, forecast verification, logarithmic score, multivariate normal distribution}
\end{abstract}

\section{Introduction}
\label{sec1}
More than 30 years ago, ensemble forecasting has been established as a discipline in numerical weather prediction \citep{lewis.2005}. Ensembles are required to  generate useful probabilistic weather forecasts which capture situation-dependent forecast uncertainties \citep{leutbecher.palmer.2008}. Users of ensemble forecasts and developers working on ensemble prediction methodologies require metrics, known as scores, that assess how ``good'' the ensemble forecasts are. A very useful subset of scores are \emph{proper scores} \citep{gneiting.raftery.2007}, which are often used as loss functions in estimation problems. This study focuses on a particular proper score, the logarithmic score \citep{good.1952}, defined as the negative logarithm of the predictive density estimated from the forecast ensemble evaluated at the corresponding observation. The score is also known as ignorance score due to its link to information theory \citep{roulston.smith.2002}. In an earlier study, \citet{siegert.ea.2019} determined how the logarithmic score depends on ensemble size in the univariate case under assumptions of normality and exchangeability of the ensemble members. Here, we extend their work to general multivariate normal distributions.

To date, the development of ensemble forecasts tends to focus on improving the skill of univariate predictions. Usually, the probabilistic skill of predicting scalars is determined for a basket of variables, vertical levels of the atmosphere, lead times and regions \citep[see e.g.][]{haiden.ea.2023,inverarity.ea.2023,mctaggart-cowan.ea.2022}. However, some applications depend on the joint prediction of several variables. \citet{schefzik.ea.2013} list air traffic control, air quality and flood management as examples, but one can also consider, for instance, scenarios of wind power generation \citep{pinson.ea.2009}. 

In multivariate verification, both forecasts \ $\boldsymbol{x}$ \ and observations \ $\boldsymbol{y}$ \ are vectors of the $p$-dimensional space \ $\mathbb{R}^p$, \ where the dimension \ $p$ \  
depends on the specific application. 
The vectors do not need to correspond to vectors that appear in physics such as wind vectors.  They can consist of the values of a scalar variable at \ $p$ \ different locations for instance \citep{schefzik.2016b}, whereas other applications may consider the evolution of a scalar variable over \ $p$ \ time steps \citep{hlk.2015,llhb.2023}, or \ $p$ \ different variables \citep{mlt.2013} or any combination of these situations \citep{schefzik.2016a}. To verify multivariate probabilistic forecasts, \citet{gneiting.ea.2008} proposes the energy score, which is a multivariate generalisation of the continuous ranked probability score, and the logarithmic score. When the predictive density is a multivariate normal distribution, the latter is equivalent to a two-moment score introduced by  \citet{dawid.sebastiani.1999} (see next section for further details). \citet{wilks.2017} compares five different methods to assess reliability of multivariate forecasts including the Gaussian Box ordinate transform (BOT) proposed by \citet{gneiting.ea.2008}. This measure of reliability is based on one of the terms appearing in the logarithmic score of a multivariate Gaussian distribution and closely linked to the reliability of ensemble covariances. Additionally, \citet{roulston.2005} introduces reliability diagrams for ensemble covariances for pairs of variables.

Estimates of probabilities from small ensembles are imprecise; 
as the ensemble size increases, the estimated probabilities become more accurate. Moreover, under certain conditions, exact statements are possible regarding the dependence of probabilistic scores on ensemble size \citep{richardson.2001,ferro.ea.2008,siegert.ea.2019,leutbecher.2018}. This permits to adjust estimates of sample means of scores computed from an $n$-member ensemble to an $m$-member ensemble. When  considering the limit \ $m\to\infty$, \ so called fair scores are obtained. The notion of fair scoring rules for ensembles was introduced by \citet{fricker.ea.2013} under restrictive assumptions of independence and reliability. \citet{ferro.2014} extends the concept of fair scores to situation of exchangeable ensemble members and drops the need for reliability. When the methodology for generating ensemble members provides a sample of exchangeable forecasts, this can be exploited in order to obtain meaningful estimates of score differences between two sets of ensemble forecasts in the large ensemble size limit from numerical experiments with small ensembles \citep{leutbecher.2018}. This has the advantage of making the development process in numerical weather prediction computationally more efficient. 

A key assumption required for the derivation of the ensemble size dependence of the logarithmic score is that the ensemble members are sampling a multivariate normal distribution. Nonetheless, climatological distributions of atmospheric weather variables can deviate considerably from normality \citep[see e.g.][]{sardeshmukh.sura.2009}. Likewise, ensemble forecasts will also exhibit various levels of deviations from normality depending on forecast lead time and considered variable. Therefore, it is of interest to explore whether the relationships for the logarithmic score  derived under assumptions of multivariate normality are  useful approximations when applied to real numerical weather prediction (NWP) ensembles. Multivariate verification examples will be studied for an operational ensemble of the European Centre for Medium-Range Weather Forecasts (ECMWF). Since July 2023, the ECMWF's extended-range forecasts consist of a daily 100-member ensemble \citep{vitart.ea.2022}. The resolution of these forecasts is TCo319, which corresponds to an average mesh size of 36\,km. The methodology will be applied to ensemble sizes from 8 to 100 and a collection of different multivariate verification scenarios up to a dimension of $p=12$. The evaluation focusses on upper air variables in the medium-range (Day 1--15) verified against analyses.

We will first derive the ensemble size dependence of the logarithmic score for forecasts issued as multivariate normal distributions in Section~\ref{sec2}. Then, the data and verification methodology are described in Section~\ref{sec3}. Subsequently, the dependence on ensemble size of the logarithmic score and the fair logarithmic score are documented for several multivariate predictands in Section~\ref{sec4}. Discussion and conclusions follow in Sections~\ref{sec5} and \ref{sec6}, respectively.

\section{Derivation}
\label{sec2}
In what follows, let \ $\boldsymbol x_1,\boldsymbol x_2, \ldots ,\boldsymbol x_n$ \ be an independent sample drawn from a $p$-dimensional Gaussian distribution \ ${\mathcal N}_p\big(\boldsymbol\mu, \boldsymbol\Sigma\big)$ \ with mean vector \ $\boldsymbol \mu$ \ and covariance matrix \ $\boldsymbol\Sigma$, \ representing an $n$-member forecast ensemble, where we assume that \ $\boldsymbol\Sigma$ \ is regular. Furthermore, denote by \ ${\boldsymbol m}$ \ and \ $\boldsymbol{\mathsf S}$ \ the sample mean vector and sample covariance matrix, respectively, that is
\begin{equation*}
  {\boldsymbol m}
  :=\frac 1n\sum_{i=1}^n \boldsymbol x_i \qquad  \text{and} \qquad \boldsymbol{\mathsf S}:=\frac 1{n-1}\sum_{i=1}^n\big(\boldsymbol x_i-{\boldsymbol m}\big)\big(\boldsymbol x_i-{\boldsymbol m}\big)^{\top}.
\end{equation*}
In general, sample mean  and sample covariance matrix are unbiased estimators of the corresponding population mean \ $\boldsymbol\mu$ \ and population covariance matrix \ $\boldsymbol{\Sigma}$; \ moreover, if the latter is regular, then for \ $n>p$ \ matrix \ $\boldsymbol{\mathsf S}$ \ is  almost surely regular as well. Note that in the Gaussian case  \ ${\boldsymbol m}$ \ and \ $\frac{n-1}n\boldsymbol{\mathsf S}$ \ are also the maximum-likelihood estimators of  \ $\boldsymbol\mu$ \ and  \ $\boldsymbol{\Sigma}$, \ respectively \citep[][Theorem 3.2.1]{anderson.2003}.

The logarithmic score of a Gaussian predictive distribution \ ${\mathcal N}_p\big(\boldsymbol\mu, \boldsymbol\Sigma\big)$ \ for the observation vector \ $\boldsymbol y$ \ is given by
\begin{equation}
  \label{eq:logs}
  \logs\big(\boldsymbol\mu,\boldsymbol\Sigma;\boldsymbol y\big)=
  \frac p2 \log (2\pi) + \frac 12 \log\big(|\boldsymbol\Sigma|\big) 
  + \frac 12 \big(\boldsymbol y - \boldsymbol\mu\big)^{\top}{\boldsymbol\Sigma}^{-1}\big(\boldsymbol y - \boldsymbol\mu\big), 
\end{equation}
where \ $|\boldsymbol A|$ \ denotes the determinant of a matrix \ $\boldsymbol A$. \ However, in our study we focus on the sample version of \eqref{eq:logs}
\begin{equation}
  \label{eq:logssamp}
  \logs\big({\boldsymbol m},\boldsymbol{\mathsf S};\boldsymbol y\big)=  \frac p2 \log (2\pi) + \frac 12 \log\big(|\boldsymbol{\mathsf S}|\big) 
  + \frac 12 \big(\boldsymbol y - {\boldsymbol m}\big)^{\top}\boldsymbol{\mathsf S}^{-1}\big(\boldsymbol y - {\boldsymbol m}\big), 
\end{equation}
which correspond to the situation when an $n$-member ensemble forecasts is transformed to a Gaussian predictive distribution with mean vector \ ${\boldsymbol m}$ \ and covariance matrix \ $\boldsymbol{\mathsf S}$.

As the following calculations will show, the sample logarithmic score \eqref{eq:logssamp} is not an unbiased estimator of the population score defined by \eqref{eq:logs} and the bias depends on the ensemble size. Our aim is to derive an ensemble size dependent fair version \ $\logs_n^F$ \ of the logarithmic score which is unbiased, that is
\begin{equation*}
  {\mathsf E}\left[\logs_n^F\big({\boldsymbol m},\boldsymbol{\mathsf S};\boldsymbol y\big)\right] =  \logs\big(\boldsymbol\mu,\boldsymbol\Sigma;\boldsymbol y\big), 
\end{equation*}
where \ $\mathsf{E\left[ \cdot \right]}$ \ denotes expectation. 
In order to derive \ $\logs_n^F$, \ first one has to calculate the expectation of the sample score \ $\logs\big({\boldsymbol m},\boldsymbol{\mathsf S};\boldsymbol y\big)$, \ where the two terms depending of the forecast ensemble and representing sharpness and reliability can be treated separately.

\subsection{Reliability}
\label{subs2.1}
Consider first the reliability term
\begin{equation*}
  \rel\big({\boldsymbol m},\boldsymbol{\mathsf S};\boldsymbol y\big):=\frac 12\big(\boldsymbol y - {\boldsymbol m}\big)^{\top}\boldsymbol{\mathsf S}^{-1}\big(\boldsymbol y - {\boldsymbol m}\big),
  \end{equation*}
  depending on the squared Mahalanobis distance of the ensemble mean and the observation vector. Note that for a Gaussian sample, the sample mean \ ${\boldsymbol m}$ \ is also Gaussian with mean vector \ $\boldsymbol\mu$ \ and covariance matrix \ $n^{-1}\boldsymbol\Sigma$, \ while \ $(n-1)\boldsymbol{\mathsf S}$ \ follows a $p$-dimensional Wishart distribution \ $\mathcal W_p(n-1,\boldsymbol\Sigma)$ \ with \ $n-1$ \ degrees of freedom and scale matrix \ $\boldsymbol\Sigma$; \ moreover, \ ${\boldsymbol m}$ \ and \ $\boldsymbol{\mathsf S}$ \ are independent \citep[see e.g.][Theorem 5.7]{hardle.simar.2019}.

A short calculation based on the independence of the sample mean and sample covariance matrix implies
  \begin{equation}
    \label{eq:rel_mean}
    {\mathsf E}\left[2\rel\big({\boldsymbol m},\boldsymbol{\mathsf S};\boldsymbol y\big)\right]
    = {\mathsf E}\left[\big({\boldsymbol m}-\boldsymbol\mu\big)^{\top}\boldsymbol{\mathsf S}^{-1}\big( {\boldsymbol m}-\boldsymbol\mu\big)\right]+\big(\boldsymbol y-\boldsymbol\mu\big)^{\top}{\mathsf E}\left[\boldsymbol{\mathsf S}^{-1}\right]\big(\boldsymbol y-\boldsymbol\mu\big). 
  \end{equation}
Now, the scaled squared Mahalanobis distance of the sample mean and the population mean
  \begin{equation*}
    T^2:=n\big({\boldsymbol m}-\boldsymbol\mu\big)^{\top}\boldsymbol{\mathsf S}^{-1}\big( {\boldsymbol m}-\boldsymbol\mu\big)
  \end{equation*}
  follows a $p$-variate Hotelling's $T^2$-distribution \ $T^2_{p,n-1}$ \ with \ $(n-1)$ \ degrees of freedom \citep[][Corollary 5.3]{hardle.simar.2019}. Moreover, the $T^2$-distribution \ $T^2_{p,n-1}$ \ is proportional to the $F$-distribution \ $F_{p,n-p}$ \ with \ $p$ \ and \ $n-p$ \ degrees of freedom \citep[][Theorem 5.9]{hardle.simar.2019},
  \begin{equation*}
    \frac{n-p}{p(n-1)}T^2 \sim F_{p,n-p},
  \end{equation*}
  which for \ $n>p+2$ \ has a mean of \ $(n-p)/(n-p-2)$. \ Hence,
  \begin{equation}
    \label{eq:mean1}
    {\mathsf E}\Big[\big({\boldsymbol m}-\boldsymbol\mu\big)^{\top}\boldsymbol{\mathsf S}^{-1} \big( {\boldsymbol m}-\boldsymbol\mu\big)\Big] =  {\mathsf E}\left[\frac 1n T^2\right] = \frac{p(n-1)}{n(n-p-2)} \qquad \text{if} \qquad n>p+2. 
   \end{equation}

Furthermore, \ $(n-1)^{-1}\boldsymbol{\mathsf S}^{-1}$ \ follows a $p$-dimensional inverted Wishart distribution \ $\mathcal W_p^{-1}\big(n-1,\boldsymbol\Sigma^{-1}\big)$ \ with \ $n-1$ \ degrees of freedom, therefore, for \ $n>p+2$ \ one has
   \begin{equation}
     \label{eq:mean2}
     {\mathsf E}\left[\boldsymbol{\mathsf S}^{-1}\right]= \frac{n-1}{n-p-2} \boldsymbol\Sigma^{-1}, 
   \end{equation}
see e.g. \citet[][Lemma 7.7.1]{anderson.2003}. Note that in the scalar case \ $(p=1)$ \ \eqref{eq:mean2} gives back the second term in the corresponding equation (A9) of \citet{siegert.ea.2019}.    

Finally, combination of \eqref{eq:mean1} and \eqref{eq:mean2} with \eqref{eq:rel_mean} results in
\begin{equation}
  \label{eq:rel}
  {\mathsf E}\left[\rel\big({\boldsymbol m},\boldsymbol{\mathsf S};\boldsymbol y\big)\right]=\frac{n-1}{2(n-p-2)} \left[\frac pn + \big(\boldsymbol y-\boldsymbol\mu\big)^{\top}\boldsymbol\Sigma^{-1} \big(\boldsymbol y-\boldsymbol\mu\big)\right]. 
\end{equation}

\subsection{Sharpness}
\label{subs2.2}
The sharpness term
\begin{equation*}
  \shp\big(\boldsymbol{\mathsf S}\big):=\frac 12 \log\big(|\boldsymbol{\mathsf S}|\big)
\end{equation*}
in \eqref{eq:logssamp} depends only on the determinant of the sample covariance matrix, which is also known as sample generalized variance. In the Gaussian case, the distribution of \ $|\boldsymbol{\mathsf S}|$ \ is proportional to the distribution of the product of \ $p$ \ independent components, where the $i$th component follows a chi-square distribution \ $\mathcal X_{n-i}^2$ \ with \ $n-i$ \ degrees of freedom \citep[see e.g.][Theorem 7.5.3]{anderson.2003}. In particular,
\begin{equation*}
  |\boldsymbol{\mathsf S}| \sim |\boldsymbol\Sigma|(n-1)^{-p}\prod_{i=1}^p \xi_i, \qquad \text{where} \qquad \xi_i \sim \mathcal X_{n-i}^2, 
\end{equation*}
$i=1,2, \ldots ,p$. \ Hence,
\begin{equation}
  \label{eq:mean_logS1}
  {\mathsf E}\left[\log\!\big(|\boldsymbol{\mathsf S}|\big)\right]\!=\!\log\!\big(|\boldsymbol\Sigma| \big) \!-\! p\log(n\!-\!1) +\! \sum_{i=1}^p {\mathsf E}\left[\log (\xi_i)\right].
\end{equation}
Now, we exploit that a chi-square distribution \ ${\mathcal X}^2_{\nu}$ \ with \ $\nu$ \ degrees of freedom is equivalent to a Gamma distribution \ $\Gamma_{\nu/2,2}$ \ with shape \ $\nu/2$ \ and scale \ $2$ \ \citep[see e.g. \ (4.54) in ][]{wilks.2019}. Thus,  \ $\log (\xi_i)$ \ follows an exponential-gamma distribution with mean
\begin{equation}
  \label{eq:log-expection-chi2}
    {\mathsf E}\left[\log (\xi_i)\right] = \psi\left(\frac {n-i}2\right) + \log(2), \quad i=1,2, \ldots p, 
\end{equation}
where \ $\psi$ \ denotes the digamma function, see \citet[][Appendix A1]{siegert.ea.2019}. 
With \eqref{eq:mean_logS1} and \eqref{eq:log-expection-chi2},
we have 
\begin{equation}
  \label{eq:shp}
  {\mathsf E}\left[\shp \big(\boldsymbol{\mathsf S}\big)\right]=\frac 12 \bigg[\log\big(|\boldsymbol\Sigma| \big) - p\log\Big(\frac {n-1}2\Big) +  \psi_p\Big(\frac {n-1}2\Big)\bigg], 
\end{equation}
with
\begin{equation}
  \label{eq:mvdigamma}
  \psi_p\left(\frac {n-1}2\right):= \sum_{i=1}^p \psi\left(\frac {n-i}2\right).
\end{equation}
Note that the function \ $\psi_p$ \ is known as the multivariate digamma function, defined as the derivative of the logarithm of the multivariate gamma function, and \eqref{eq:mvdigamma} is a direct consequence of the representation of the latter given by \citet[][Theorem 1.4.6]{gupta.nagar.1999}. 
  
\subsection{The fair logarithmic score}
\label{subs2.3}
Combining \eqref{eq:logssamp} with \eqref{eq:rel} and \eqref{eq:shp}, one gets
\begin{align}
  \nonumber
  {\mathsf E\left[\logs\big({\boldsymbol m},\boldsymbol{\mathsf S};\boldsymbol y\big)\right] =}&\, \frac p2 \log (2\pi) + \frac 12 \log\big(|\boldsymbol\Sigma| \big) \\ \label{eq:logsmean} &+ \frac 12  \frac{n-1}{n-p-2}\big(\boldsymbol y-\boldsymbol\mu\big)^{\top}\boldsymbol\Sigma^{-1}\big(\boldsymbol y-\boldsymbol\mu\big) \\ & + \frac 12 \left[ \psi_p\Big(\frac {n\!-\!1}2\Big) - p\log\Big(\frac {n\!-\!1}2\Big) +  \frac{p(n-1)}{n(n-p-2)}\right]. \nonumber 
\end{align}
Now, an unbiased estimator of the population logarithmic score can be obtained by multiplying the reliability term in \eqref{eq:logssamp} with \ $(n-p-2)/(n-1)$ \ and subtracting the remaining bias, which is then independent
of the population parameters \ $\boldsymbol\mu$ \ and \ $\boldsymbol\Sigma$. \ This approach results in the following formula of the fair logarithmic score
\begin{align}
  \nonumber
  \logs_n^F\big({\boldsymbol m},\boldsymbol{\mathsf S};\boldsymbol y\big)=&\,  \frac p2 \log (2\pi) + \frac 12 \log\big(|\boldsymbol{\mathsf S}|\big) \\
            \label{eq:fairlogs} &+ \frac{n-p-2}{2(n-1)}\big(\boldsymbol y - {\boldsymbol m}\big)^{\top}\boldsymbol{\mathsf S}^{-1}\big(\boldsymbol y - {\boldsymbol m}\big) \\
  & - \frac 12 \left[ \psi_p\Big(\frac {n-1}2\Big) - p\log\Big(\frac {n-1}2\Big) +  \frac pn\right]. \nonumber
\end{align}
Note, that \eqref{eq:fairlogs} is a direct multivariate generalization of the corresponding formula (13) of \citet{siegert.ea.2019}.

\subsection{The ensemble-adjusted logarithmic score}
\label{subs2.4}
Now, we assume that we have an $n$-member and an $N$-member ensemble forecast \ $(n\ne N)$ \  following the same Gaussian law and having ensemble mean vectors \ $\boldsymbol m_n$ \ and  \ $\boldsymbol m_N$ \ and ensemble covariance matrices \ $\boldsymbol{\mathsf S}_n$ \ and \ $\boldsymbol{\mathsf S}_N$, \ respectively. According to \eqref{eq:logsmean},
\begin{equation*}
  {\mathsf E}\left[\logs\big({\boldsymbol m}_n,\boldsymbol{\mathsf S}_n;\boldsymbol y\big)\right]\ne {\mathsf E}\left[\logs\big({\boldsymbol m}_N,\boldsymbol{\mathsf S}_N;\boldsymbol y\big)\right].
\end{equation*}
The ensemble-adjusted logarithmic score \ $\logs_{n \to N}^{EA}$ \ uses the mean vector and the covariance matrix estimated from the $n$-member ensemble; however, it has the same expectation as the sample logarithmic score based on the $N$-member ensemble, see also \citet[][Section 2]{siegert.ea.2019}.
The same arguments as in the case of the fair logarithmic score lead us to
\begin{align*}
  \label{eq:ealogs}
  \nonumber
  \logs_{n \to N}^{EA}\big({\boldsymbol m}_n,\boldsymbol{\mathsf S}_n;\boldsymbol y\big) =&\, \frac p2 \log (2\pi) + \frac 12 \log\big(|\boldsymbol{\mathsf S}_n|\big) + \frac{p(N-1)(n-N)}{2nN(N-p-2)} \\ \nonumber
 &+ \frac{N-1}{N-p-2}\frac{n-p-2}{2(n-1)}\big(\boldsymbol y - {\boldsymbol m}_n\big)^{\top}\boldsymbol{\mathsf S}^{-1}_n\big(\boldsymbol y - {\boldsymbol m}_n\big) \\ 
 &- \frac 12 \left[ \psi_p\Big(\frac {N\!-\!1}2\Big) -  \psi_p\Big(\frac {n-1}2\Big) + p\log\Big(\frac {n-1}{N-1}\Big)\right], \nonumber
  \end{align*}
  which generalizes the corresponding formula (11) of \citet{siegert.ea.2019}. Using again  \eqref{eq:rel} and \eqref{eq:shp}, one can easily verify the required equality
\begin{equation*}
  {\mathsf E}\left[\logs_{n \to N}^{EA}\big({\boldsymbol m}_n,\boldsymbol{\mathsf S}_n;\boldsymbol y\big)\right] = {\mathsf E}\left[\logs\big({\boldsymbol m}_N,\boldsymbol{\mathsf S}_N;\boldsymbol y\big)\right].
\end{equation*}
Furthermore, short straightforward calculation using \ \begin{equation*}
  \lim_{x\to\infty}\big[\psi_p(x) - p\log(x)\big]=0
\end{equation*}
shows
\begin{equation*}
  \lim_{N\to\infty}\logs_{n \to N}^{EA}\big({\boldsymbol m}_n,\boldsymbol{\mathsf S}_n;\boldsymbol y\big) =  \logs_n^F\big({\boldsymbol m}_n,\boldsymbol{\mathsf S}_n;\boldsymbol y\big),
\end{equation*}
that is for an infinitely large ensemble the ensemble adjusted logarithmic score equals the fair one.

\subsection{Ensemble size dependence for reliable ensembles}
\label{subs2.5}
\begin{figure*}[t]
    \centering
    \includegraphics[width=0.8\linewidth]{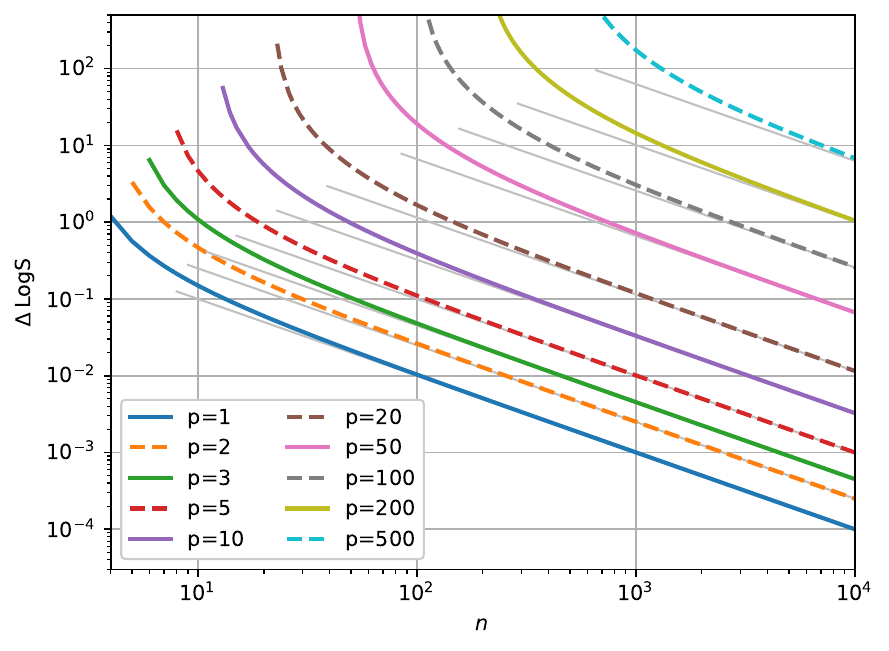}
    \caption{Ensemble size dependence of \ $\Delta\logs$ \ for reliable $n$-member ensemble forecasts of $p$-dimensional vectors. The straight sloping lines show the asymptotic  behaviour for large ensemble size given by \ $\frac{p(p+3)}{4n}$.}
    \label{fig:delta-logs}
\end{figure*}

Now, we calculate the expected logarithmic score for reliable $n$-member ensembles. Members of a reliable ensemble sample the same $p$-variate multivariate distribution \ ${\mathcal N}_p\big(\boldsymbol\mu, \boldsymbol\Sigma\big)$ \ as the observation. The expected squared Mahanalobis distance of the observation from the distribution mean is given by
\begin{equation}
\label{eq:mahanalobis-obs-dist-mean}
  {\mathsf E}_{\boldsymbol y}\big(\boldsymbol y\!-\!\boldsymbol\mu\big)^{\top}\boldsymbol\Sigma^{-1}\big(\boldsymbol y\!-\!\boldsymbol\mu\big) = \mathrm{tr}\left(\boldsymbol\Sigma^{-1}\boldsymbol\Sigma\right)=p,
\end{equation}
where \ $\mathrm{tr}\left(\boldsymbol{\mathsf{A}}\right)$ \ denotes the trace of a matrix \ $\boldsymbol{\mathsf{A}}$. \ The expectation with respect to the observation of equations
\eqref{eq:logsmean} and \eqref{eq:logs} yields
\begin{align*} 
  {\mathsf E}_{\boldsymbol y}{\mathsf E}\left[\logs\big({\boldsymbol m},\boldsymbol{\mathsf S};\boldsymbol y\big)\right] =&\, \frac p2 \log (2\pi)
  + \frac 12 \log\big(|\boldsymbol\Sigma| \big) 
   + \frac p2  \frac{n-1}{n-p-2}  \\ &+ \frac 12 \bigg[ \psi_p\Big(\frac {n\!-\!1}2\Big) - p\log\Big(\frac {n\!-\!1}2\Big) +  \frac{p(n\!-\!1)}{n(n\!-\!p\!-\!2)}\bigg],
\\[2mm]
 {\mathsf E}_{\boldsymbol y} \logs\big(\boldsymbol\mu,\boldsymbol\Sigma;\boldsymbol y\big)=&\,
  \frac p2 \log (2\pi) + \frac 12 \log\big(|\boldsymbol\Sigma|\big) + \frac p2, 
\end{align*}
where  \eqref{eq:mahanalobis-obs-dist-mean} has been used to get the expected squared Mahanalobis distance. The difference of the two preceding equations is 
\begin{align}
  \label{eq:logs-reliable-ens}
\Delta\logs :=&\,
   {\mathsf E}_{\boldsymbol y}{\mathsf E}\left[\logs\big({\boldsymbol m},\boldsymbol{\mathsf S};\boldsymbol y\big)\right] - 
   {\mathsf E}_{\boldsymbol y} \logs\big(\boldsymbol\mu,\boldsymbol\Sigma;\boldsymbol y\big) \\ =&\,\frac p2 \,\frac{np\!+\!2n\!-\!1}{n(n\!-\!p\!-\!2)} 
   +\frac 12 \left[ \psi_p\Big(\frac {n\!-\!1}2\Big) - p\log\Big(\frac {n\!-\!1}2\Big)  \right]  \nonumber
\end{align}
Thus, the difference between the expected logarithmic score of a reliable $n$-member ensemble and the expected logarithmic score of a reliable Gaussian distribution is a universal function of \ $p$ \ and \ $n$ \ and does not depend on the covariance matrix.
Figure~\ref{fig:delta-logs} shows this difference in the expected logarithmic score for ensemble sizes in the range \ $p+3$ \ to $10^4$ and vectors with dimensions in the range 1 to 500.

An asymptotic relationship can be derived for large ensemble size by using an approximation of the Digamma function
for large real arguments \cite[see 6.3.18 in][]{abramowitz.stegun.1964}
\begin{equation*}
  \psi(z) = \log (z) \textstyle
  -\frac{1}{2z}  + O\big(z^{-2}\big).
\label{eq:psi-asym}
\end{equation*}
This implies that the multivariate Digamma function can be approximated as
\begin{equation*}
    \psi_p(z) = p\log(z) - \frac{p(p+1)}{2z} + O(z^{-2}).
\end{equation*}
Inserting this in \eqref{eq:logs-reliable-ens} leads to the following asymptotic relationship for 
\begin{equation}
    \Delta\,\logs =\frac{p(p+3)}{4n} + O\big(n^{-2}\big)
   \label{eq:logs-reliable-asymptotic}
\end{equation}
for large ensemble size \ $n$. \ Approximation \eqref{eq:logs-reliable-asymptotic} is  very accurate for sufficiently large ensemble size (Fig.~\ref{fig:delta-logs}); however, the minimum number of ensemble members needed for the asymptotic relationship to be sufficiently precise increases with increasing dimension \ $p$ \ of the predicted vector. 

\section{Data and methodology}
\label{sec3}
This section introduces the NWP data and the verification methodology. Multiple multivariate prediction scenarios are considered at lead times of up to 15~days using operational ensemble forecasts from ECMWF. 

\subsection{Forecasts and analyses}
\label{subs3.1}
In order to consider a large range of ensemble sizes, it was decided to use data from the extended-range ensemble, which runs daily with 100 perturbed members since July 2023 \citep{vitart.ea.2022}. The horizontal resolution of the forecast is about 36\,km (TCo319, triangular truncation at horizontal wavenumber 319 with a cubic octahedral grid). Apart from horizontal resolution, time step and the larger ensemble size, the configuration of the extended-range ensemble is identical to the configuration of the medium-range ensemble. \citet{lang.ea.2021b} describe a recent version of the ensemble methodology used with ECMWF's Integrated Forecasting System (IFS), where the version introduced in July 2023 is known as cycle 48r1 \citep{lang_rodwell_schepers.2023}.

\begin{table*}[t]
 \caption{Configurations of $p$-dim vector predictands. The variables (var) $T$, ws, $z$, $u,v$ refer to temperature, wind speed, geopotential and the zonal and meridional wind component, respectively.}
 \centering
 \medskip
 \begin{tabular}{rcclcp{0.22\linewidth}}
   %
   %
    \hline
  $p$ & type & var   & level(s) (hPa) &  stencil  & configuration \\ 
    \hline
    4 & stencil & $T$   & 850                   &  $2\times2$   & T850dxdyL4 \\
    9 & stencil & $T$   & 850                   &  $3\times3$   & T850dxdyL9 \\    
    4 & stencil & ws    & 850                   &  $2\times2$   & ws850dxdyL4 \\
    9 & stencil & ws    & 850                   &  $3\times3$   & ws850dxdyL9 \\    
    4 & stencil & $z$   & 500                   &  $2\times2$   & z500dxdyL4 \\
    9 & stencil & $z$   & 500                   &  $3\times3$   & z500dxdyL9 \\    
    3 & profile & $z$   & 200, 500, 925         &    $1$   & z200to925L3 \\
    6 & profile & $z$   & 200, \dots, 925       &    $1$   & z200to925L6 \\
    2 & vector wind & $u,v$ & 850                   &    $1$   & uv850L2\\
    2 & vector wind & $u,v$ & 500                   &    $1$   & uv500L2\\
    2 & vector wind & $u,v$ & 200                   &    $1$   & uv200L2\\
    4 & vec.~wind profile & $u,v$ & 200, 850      &    $1$   & uv200to850L4\\
   12 & vec.~wind profile & $u,v$ & 200, \dots, 925 &    $1$   & uv200to925L12\\
    \hline
  \end{tabular}
  \label{tab:configs}
\end{table*}

The ECMWF ensemble forecasts can be considered as exchangeable with a few exceptions described below. The generation mechanism for the ensemble ensures that all members sample initial uncertainties and model uncertainties from the same underlying distribution. The atmospheric initial conditions are obtained by adding perturbations to an unperturbed analysis, that is the best estimate of the initial conditions. There are 50 different EDA perturbations, one each for the first 50 ensemble forecasts. The remaining forecasts reuse the same EDA perturbations, that is members $j$ and $50+j$ are constructed with the $j$-th EDA perturbation. There are 5 different initial conditions for the ocean and sea ice, which are reused every 5 members. All remaining perturbations are independent realisations for the 100 members of the extended-range forecasts. These consist of additional perturbations of the atmospheric initial conditions based on singular vectors and  the representation of model uncertainties with the Stochastically Perturbed Parametrisations scheme SPPT.

The forecasts are verified against operational ECMWF analyses. We consider the boreal autumn season with forecasts starting at 00~UTC in the period 1 September -- 30 November 2023, which provides a sample of 91 ensemble forecasts in total.

\subsection{Verification}
\label{subs3.2}

Scores are computed for vector predictands at lead times of $t=1, 2, \ldots, 15$\,d, valid at 00~UTC, where the predictands are assembled from fields on a regular 1.5\,deg $\times$ 1.5\,deg latitude/longitude grid. The spectral forecast and analyses fields are truncated at horizontal wavenumber 120 prior to the transformation on the grid. We will focus on spatially aggregated results for the northern mid-latitudes (35\degree\,N -- 65\degree\,N). Cosine latitude weights are used in the computation of the spatial averages to account for the area grid points represent.

Thirteen different vectors predictands with dimension $p$ ranging from 2 to 12 are considered (Table~\ref{tab:configs}). First we look at predictands consisting of single variables on a set of different locations with a fixed spatial separation. Let us call such a group of points a stencil. Six configurations consist of horizontally distributed points on  $2\times 2$ and $3\times 3$ stencils leading to 4-dim and 9-dim vectors. The grid points are obtained through meridional and longitudinal shifts by approximately 1000\,km. 
Nearest grid points are selected on the regular lat/lon grid for these geometries. A stencil is associated with every grid point of the lat/lon grid and the scores for that point refer to that stencil.
For the $2\times 2$ and $3\times 3$ stencils, scores have been computed for temperature on the 850\,hPa level, wind speed on the 850\,hPa level and 500\,hPa geopotential. These six configurations are referred to as T850dxdyL4, T850dxdyL9, ws850dxdyL4, ws850dxdyL9, z500dxdyL4 and z500dxdyL9, respectively. 

Two configurations consist of profiles; these are predictands in a column consisting of vertically seperated fields on 3 and on 6 pressure levels. The predictands  are geopotential on the 200, 500, 925\,hPa levels and on 200, 300, 500, 700, 850, 925\,hPa levels. These 3-dim and 6-dim configurations are referred to as z200to925L3 and z200to925L6, respectively. 

The following three 2-dim configurations consist of horizontal vector wind at three different pressure levels and are referred to as uv200L2, uv500L2 and uv850L2, respectively. The final two configurations consist of profiles of vector wind on two (six) pressure levels leading to the 4-dim  (12-dim) configurations uv200to850L4 and uv200to925L12.

Scores are computed for ensemble sizes of $n=8, 12, 16, 24, 32, 64$ and $100$ members. Computations for small ensembles are omitted for configurations with dimension $p > n-3$ as this violates the assumptions in the derivation of the fair logarithmic score.

\subsection{Testing multivariate normality}
\label{subs3.3}
Both the fair and the ensemble-adjusted logarithmic scores introduced in Sections \ref{subs2.3} and \ref{subs2.4}, respectively, are derived under the assumption of multivariate normality, and the same applies to the asymptotic formulae of Section \ref{subs2.5}. To investigate whether the forecast vectors studied in Section \ref{sec4} follow a multivariate Gaussian law, in accordance with the recommendations of \citet{mecklin.mundfrom.2005} based on extended Monte Carlo simulations, we consider the Henze-Zinkler (HZ) test \citep{henze.zirkler.1990}. Using the notations of Section \ref{sec2}, for a $p$-dimensional sample \ $\boldsymbol x_1,\boldsymbol x_2, \ldots ,\boldsymbol x_n$ \ with sample mean vector \ ${\boldsymbol m}$ \ and biased sample covariance matrix \ $\boldsymbol{\mathsf S}_n := \frac{n-1}{n}\mathsf{S}$, \ the Henze-Zirkler test statistics is defined as
\begin{equation*}
  \label{HZtest}
  T_{n,\beta}(p) :=  \frac 1n \sum_{i=1}^n\sum_{j=1}^n{\mathrm e}^{-\frac{\beta^2}2 D_{ij}}  + n\big(1+2\beta^2\big)^{-\frac p2}- 2\big(1+\beta^2\big)^{-\frac p2} \sum_{i=1}^n{\mathrm e}^{-\frac{\beta^2}{2(1+\beta^2)}D_i}, \nonumber
\end{equation*}
where
\begin{equation*}
D_{ij}:=(\boldsymbol x_i - \boldsymbol x_j)^{\top}\boldsymbol{\mathsf S}_n^{-1}(\boldsymbol x_i - \boldsymbol x_j), \qquad  
D_{i}:=(\boldsymbol x_i - \boldsymbol m)^{\top}\boldsymbol{\mathsf S}_n^{-1}(\boldsymbol x_i - \boldsymbol m),
\end{equation*}
and, following the suggestions of \citet{henze.zirkler.1990}, smoothing parameter \ $\beta>0$ \ is chosen as 
\begin{equation}
 \label{eq:beta}
 \beta= \frac 1{\sqrt{2}}\left(\frac{n(2p+1)}4\right)^{\frac 1{p+4}}.
\end{equation}
In fact, \ $T_{n,\beta}(p)$ \ is proportional to the ${\mathsf L}^2$ distance between the kernel density estimator applied to the standardized sample with a $p$-dimensional standard Gaussian kernel and the Gaussian distribution expected under the null hypothesis. The smoothing parameter specified in \eqref{eq:beta} corresponds to the optimal bandwidth of the kernel density estimator \citep[][Section 4.2]{silverman.1986}.

\begin{figure*}[t]
\makebox[0.5\linewidth]{(a)}\makebox[0.5\linewidth]{(b)}
\includegraphics[width=0.49\linewidth]{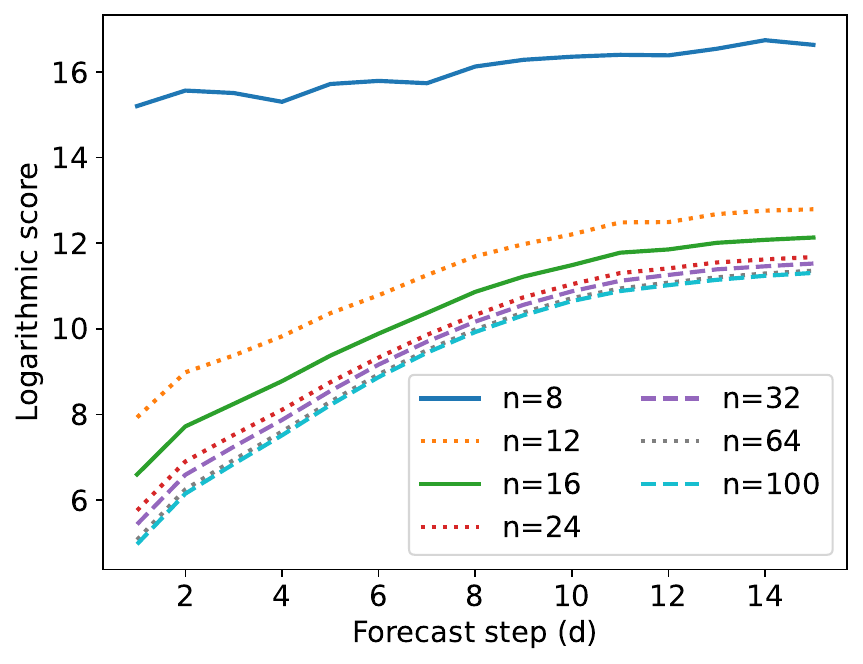}%
\includegraphics[width=0.49\linewidth]{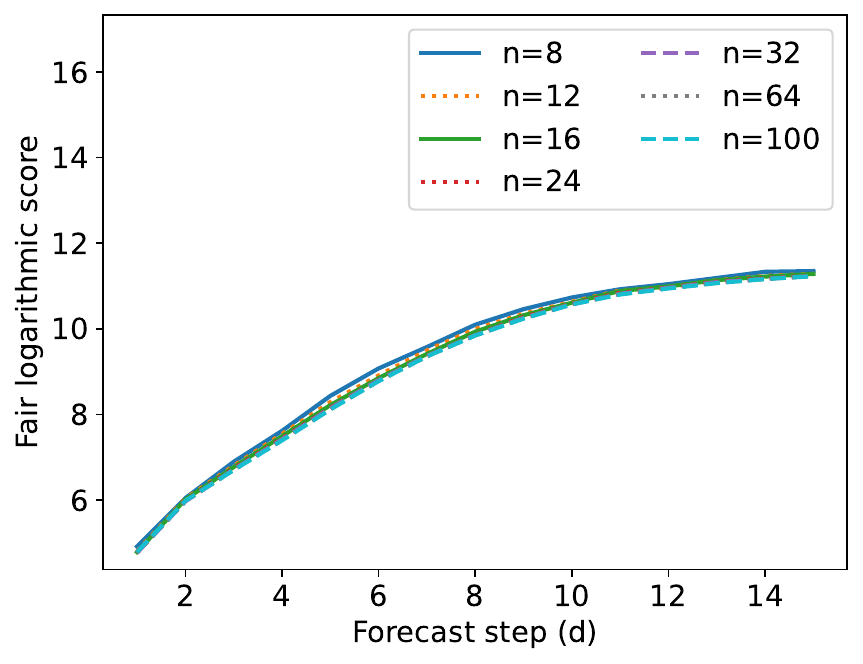}
  
  \caption{Ensemble size dependence of (a) logarithmic score and (b) fair logarithmic score for the 4-dim predictand of 850\,hPa temperature on the $2\times 2$ stencil of points (configuration  t850dxdyL4). The ensemble size ranges from $n=8$ to $n=100$.}
\label{fig:t850dxdyL4}
\end{figure*}

If the sample is drawn from a multivariate Gaussian distribution, then the limiting distribution of \ $T_{n,\beta}(p)$ \ as \ $n\to\infty$ \ is approximately log-normal with mean \ $\mathfrak m_{\beta,p}$ \ and variance \ $\mathfrak s^2_{\beta,p}$, \ where
\begin{align*}
  \mathfrak m_{\beta,p}&= 1-\big(1+2\beta^2\big)^{-\frac p2}\left(1+\frac{p\beta^2}{1+2\beta^2}+\frac{p(p+2)\beta^4}{2(1+2\beta^2)^2} \right), \\
  \mathfrak s^2_{\beta,p} &= 2\big(1+4\beta^2\big)^{-\frac p2} +2\big(1+2\beta^2\big)^{-p} \left(1+ \frac{2p\beta^4}{(1+2\beta^2)^2} + \frac{3p(p+2)\beta^8}{4(1+2\beta^2)^4}\right) \\
&\quad -4 
\omega^{-\frac p2}\left(1+\frac{3p\beta^4}{2\omega}+\frac{p(p+2)\beta^8}{2\omega^2}\right),
\end{align*}
with \ 
$\omega :=\big(1+\beta^2\big)\big(1+3\beta^2\big)$. 
\ Hence, \ $\log\big(T_{n,\beta}(p)\big)$ \ is approximately Gaussian with mean 
\ $\mu_{\beta,p}$ \ and variance \ $\sigma^2_{\beta,p}$ \ given as
\begin{equation*}
  \mu_{\beta,p} =\log\left(\frac{ \mathfrak m_{\beta,p}^2}{\sqrt{ \mathfrak s_{\beta,p}^2+ \mathfrak m_{\beta,p}^2} }\right) \qquad \text{and} \qquad \sigma^2_{\beta,p}=\log\left(1+\frac{ \mathfrak s_{\beta,p}^2}{\mathfrak m_{\beta,p}^2}\right). 
\end{equation*}
Thus, the significance of multivariate normality can be tested using the Wald test statistic
\begin{equation}
  \label{wald}
  Z:=\frac{\log(T_{n,\beta}(p))-\mu_{\beta,p}}{\sigma_{\beta,p}},
\end{equation}
which for large sample sizes \ $n$ \ is approximately standard Gaussian.

\section{Results}
\label{sec4}
In the following section we discuss the behaviour of the logarithmic score and its fair version for the various multivariate predictands introduced in Section \ref{subs3.2}.  First, the results for predictands consisting of stencils of points are described, which is followed by forecasts corresponding to profiles. Finally, our findings for vector wind and profiles of vector wind are reported.

\begin{figure*}[t]
\makebox[0.5\linewidth]{(a)}\makebox[0.5\linewidth]{(b)}
\includegraphics[width=0.49\linewidth]{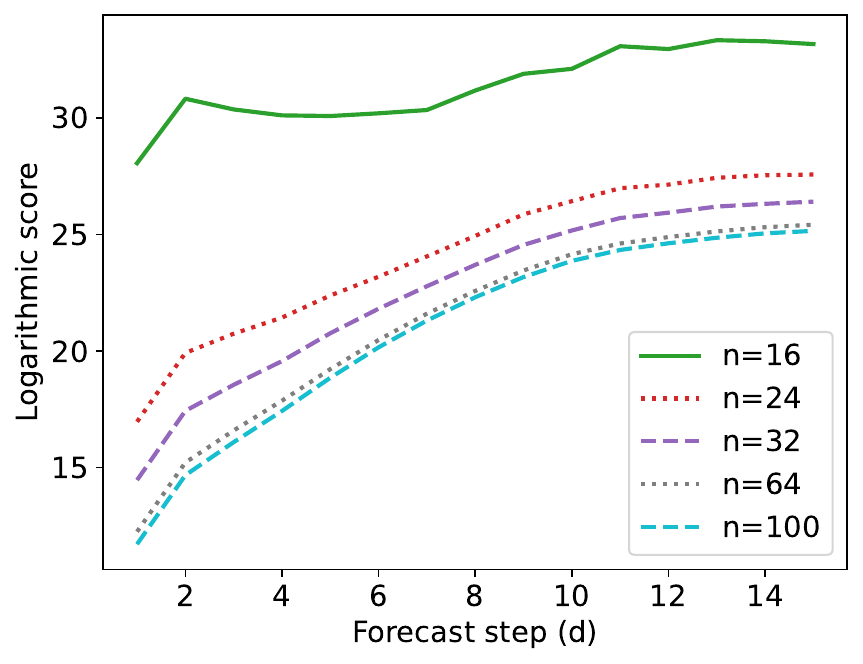}%
\includegraphics[width=0.49\linewidth]{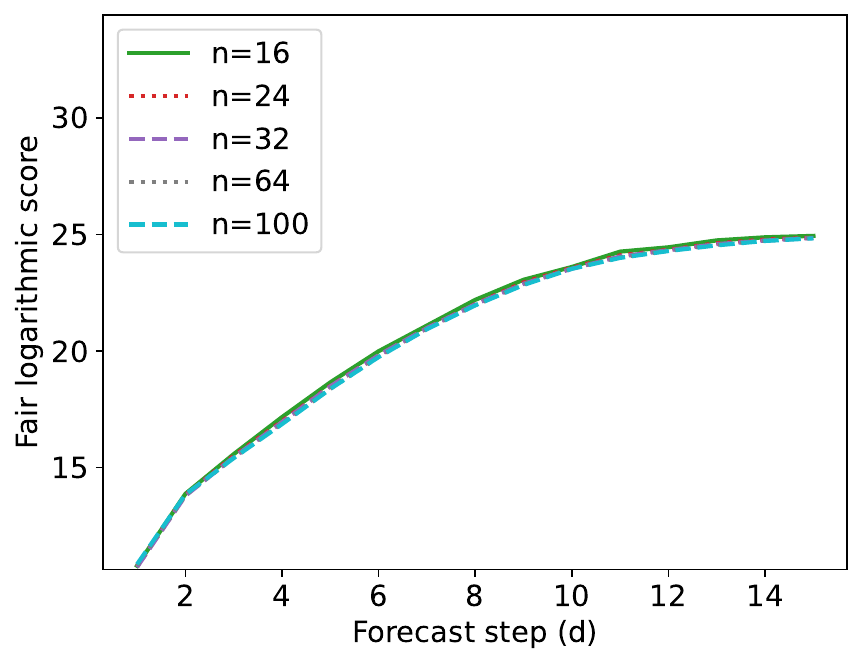}
  
  \caption{Ensemble size dependence of (a) logarithmic score and (b) fair logarithmic score for the 9-dim predictand of 850\,hPa temperature on the $3\times 3$ stencil of points (configuration  t850dxdyL9). The ensemble size ranges from $n=16$ to $n=100$.}
\label{fig:t850dxdyL9}
\end{figure*}

\subsection{Stencils}
\label{subs4.1}
Figure~\ref{fig:t850dxdyL4} shows the logarithmic score for 850\,hPa
temperature verified on the $2\times 2$ stencil. This corresponds to a
4-dim predictand of 4 points separated horizontally by 1000\,km. The
logarithmic score decreases considerably with ensemble size
(Fig.~\ref{fig:t850dxdyL4}a). For instance, the 6-day prediction with
100 members has a smaller logarithmic score than the 2-day prediction
with 12 members. In contrast, the fair logarithmic scores for the
different ensemble sizes are very similar (Fig.~\ref{fig:t850dxdyL4}b). At early lead times up to 2~days, the score values are practically identical, whereas at longer lead times the values of the fair logarithmic score for the smallest ensemble sizes (8 and 12 members) tend to be slightly larger than the 
score values for the larger ensemble sizes.

\begin{table*}[!ht]
  \centering
  \caption{Difference \ $(\Delta\mathrm{LogS})$ \  between the logarithmic scores of the 8-member ensemble and the 100-member ensemble, difference \ $(\Delta\mathrm{fLogS})$ \ between the fair logarithmic scores for the two ensemble sizes and ratio \ $(R:=\Delta\mathrm{fLogS}/\Delta\mathrm{LogS})$ \ of the score differences at lead times of 48\,h and 120\,h. All configurations of verification vectors with \ $p+3 \le 8$ \ are shown.}

\medskip  
\begin{tabular}{lrrrrrr}
\hline
              & \multicolumn{3}{c}{Day~2} & \multicolumn{3}{c}{Day~5} \\\cline{2-7}
  configuration & $\Delta\mathrm{LogS}$ & $\Delta\mathrm{fLogS}$ & $R$
  & $\Delta\mathrm{LogS}$ & $\Delta\mathrm{fLogS}$ & $R$ \\
\hline
t850dxdyL4 &   9.42 &   0.063&   0.01 &   7.50 &   0.318&   0.04 \\
ws850dxdyL4 &   7.60 &   0.040&   0.01 &   6.48 &   0.159&   0.02 \\
z500dxdyL4 &   5.24 &   0.006&   0.00 &   6.06 &   0.121&   0.02 \\
z200to925L3 &   2.54 &   0.034&   0.01 &   2.48 &   0.103&   0.04 \\
uv200L2 &   0.84 &   0.022&   0.03 &   0.89 &   0.051&   0.06 \\
uv500L2 &   0.96 &   0.023&   0.02 &   0.91 &   0.055&   0.06 \\
uv850L2 &   1.19 &   0.033&   0.03 &   1.05 &   0.088&   0.08 \\
uv200to850L4 &   7.21 &   0.039&   0.01 &   6.86 &   0.210&   0.03 \\
\hline
\end{tabular}  
  \label{tab:deltas8to100}
\end{table*}

In Figure~\ref{fig:t850dxdyL9}  the corresponding results for
850\,hPa for the $3\times 3$ stencil are given. Again, adjacent points in the stencil are separated horizontally by 1000\,km but the stencil now covers a larger area than the $2\times 2$ stencil. The dependence of the
logarithmic score on ensemble size for this 9-dim predictand is more
pronounced than for the 4-dim predictand of the $2\times 2$ stencil
(compare Figs.~\ref{fig:t850dxdyL9}a and \ref{fig:t850dxdyL4}a), while the
fair logarithmic score again exhibits only a tiny dependence on ensemble
size (Fig.~\ref{fig:t850dxdyL9}b). Note that results for the 12-member ensemble
have been omitted from the figure to avoid an excessive increase in
the range of the y-axis as the logarithmic score reaches values of
100. However, the adjustment of the fair score still works very well
for the 12-member ensemble even though the difference in the
logarithmic score between 12 and 100 members is enormous.

\begin{table*}[t]
  \centering
  \caption{Difference \ $(\Delta\mathrm{LogS})$ \ between the logarithmic scores of the 16-member ensemble and the 100-member ensemble, difference \ $(\Delta\mathrm{fLogS})$ \ between the fair logarithmic scores for the two ensemble sizes and ratio \ $(R:=\Delta\mathrm{fLogS}/\Delta\mathrm{LogS})$ \ of the score differences at lead times of 48\,h and 120\,h.  All 13 configurations of verification vectors are shown.}

\medskip
  
\begin{tabular}{lrrrrrr}
\hline
  & \multicolumn{3}{c}{Day~2} & \multicolumn{3}{c}{Day~5} \\
  \cline{2-7}
  configuration & $\Delta\mathrm{LogS}$ & $\Delta\mathrm{fLogS}$ & $R$
  & $\Delta\mathrm{LogS}$ & $\Delta\mathrm{fLogS}$ & $R$ \\
\hline
t850dxdyL4 &   1.57 &   0.033&   0.02 &   1.15 &   0.102&   0.09 \\
t850dxdyL9 &  16.15 &   0.025&   0.00 &  11.23 &   0.284&   0.03 \\
ws850dxdyL4 &   1.21 &  $-$0.001&  $-$0.00 &   1.02 &   0.065&   0.06 \\
ws850dxdyL9 &  12.75 &   0.005&   0.00 &  10.04 &   0.192&   0.02 \\
z500dxdyL4 &   0.77 &  $-$0.019&  $-$0.03 &   0.92 &   0.036&   0.04 \\
z500dxdyL9 &   9.63 &  $-$0.033&  $-$0.00 &   9.87 &   0.174&   0.02 \\
z200to925L3 &   0.56 &   0.009&   0.02 &   0.55 &   0.043&   0.08 \\
z200to925L6 &   4.15 &   0.098&   0.02 &   3.26 &   0.252&   0.08 \\
uv200L2 &   0.23 &   0.006&   0.03 &   0.25 &   0.019&   0.08 \\
uv500L2 &   0.26 &   0.002&   0.01 &   0.26 &   0.022&   0.09 \\
uv850L2 &   0.34 &   0.009&   0.03 &   0.30 &   0.034&   0.12 \\
uv200to850L4 &   1.17 &   0.026&   0.02 &   1.06 &   0.069&   0.07 \\
uv200to925L12 &  68.83 &   0.202&   0.00 &  55.09 &   0.663&   0.01 \\
\hline
\end{tabular}  
  \label{tab:deltas16to100}
\end{table*}

In addition to 850\,hPa temperature, 850\,hPa wind speed and 500\,hPa
geopotential have been verified for the stencil predictands. Table~\ref{tab:deltas8to100} lists differences in
logarithmic scores between 8 and 100 members at Days~2 and 5 for the
$2\times 2$ stencils, together with the ratio \ $R$ \ of the difference of the fair logarithmic scores to the corresponding difference of the logarithmic scores. This ratio quantifies how well the ensemble size adjustment in the fair scores works. Results for wind speed and geopotential are consistent with the results for 850\,hPa temperature: differences in the logarithmic scores are between 5 and 10, while the deviations in the fair logarithmic score do not exceed 0.3. Table~\ref{tab:deltas16to100} contains the score differences
between 16 and 100 members and the corresponding ratios \ $R$ \ for both the $2\times 2$ and the $3\times 3$ stencils and the three variables. The logarithmic score differences for the $3\times 3$ stencils are about an order of magnitude larger than the score differences for the $2\times 2$ stencils.  
Furthermore, the \ $R$ \ values suggest that the ensemble size adjustment works at least as well for the 9-dim predictands of the $3\times 3$ stencil as the 4-dim predictands of the $2\times 2$ stencils. The ratio reaches its largest value of 0.09 for $2\times 2$ stencils of temperature at 850\,hPa at Day 5, and, in general, the value of \ $R$ \ tends to be larger at Day~5 than at Day~2. In terms of the three variables, the adjustment works best for geopotential, followed by wind speed and then temperature.

\subsection{Profiles}
\label{subs4.2}

\begin{figure*}[t]
\makebox[0.5\linewidth]{(a)}\makebox[0.5\linewidth]{(b)}
\includegraphics[width=0.49\linewidth]{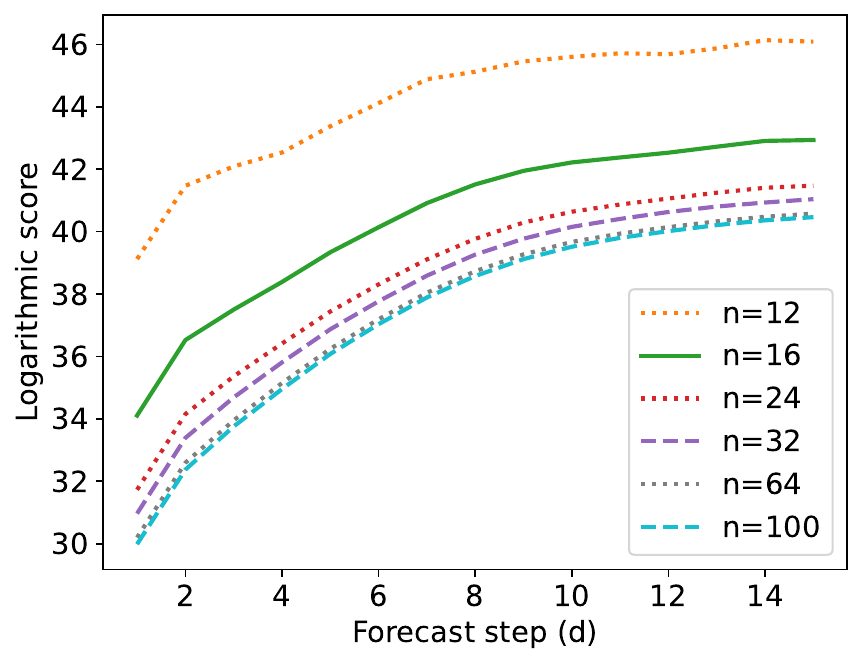}%
\includegraphics[width=0.49\linewidth]{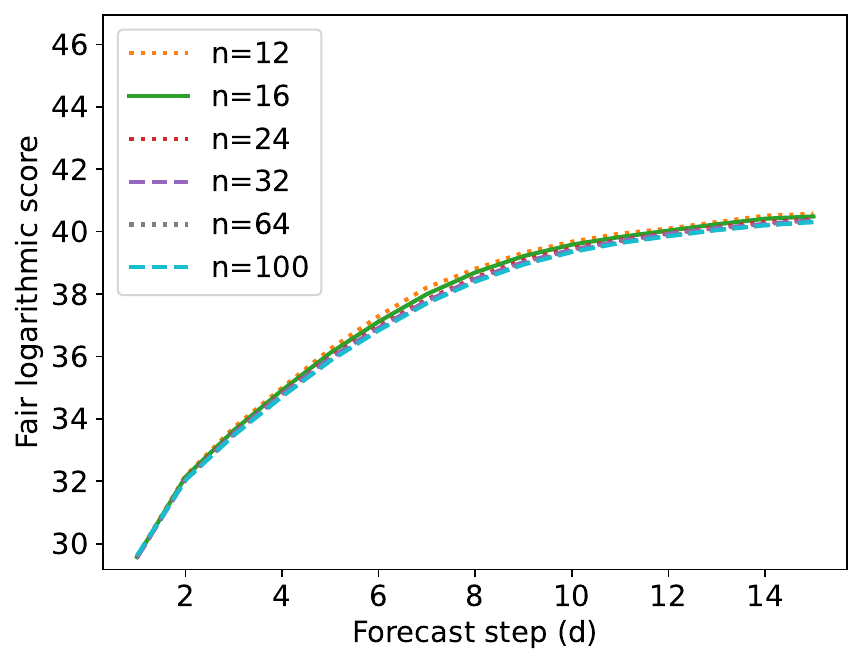}
  
  \caption{Ensemble size dependence of (a) logarithmic score and (b) fair logarithmic score for the 6-dim predictand of geopotential on the levels of 200, 300, 500, 700, 850, 925\,hPa  (configuration  z200to925L6). The ensemble size ranges from $n=12$ to $n=100$.}
\label{fig:z200to925L6}
\end{figure*}

Now, we consider profiles, i.e.\ predictands consisting of a single
variable at multiple levels.  Figure~\ref{fig:z200to925L6} shows the
logarithmic score of geopotential at six pressure levels between
200\,hPa and 925\,hPa. Again, the logarithmic score has a pronounced
sensitivity to ensemble size. The 2-day prediction with 16 members has
a larger logarithmic score than the 5-day prediction with 100 members
(Fig.~\ref{fig:z200to925L6}a). The adjustment in the fair logarithmic
score removes most of the sensitivity to ensemble size
(Fig.~\ref{fig:z200to925L6}b). The adjustment works best for early
lead times, which is consistent with the results for the stencils.

Tables~\ref{tab:deltas8to100} and \ref{tab:deltas16to100} list results
for a 3-dim predictand composed of geopotential at 200, 500 and
925\,hPa under configuration z200to925L3. The latter table also
contains the results for the 6-dim configuration z200to925L6 shown in
Figure~\ref{fig:z200to925L6}. The results imply that the ensemble size
adjustments for the geopotential profiles work equally well for the
3-dim and 6-dim predictands.

\subsection{Vector wind}
\label{subs4.3}
Next, 2-dim predictands consisting of the horizontal wind components
$u$ and $v$ at a single location are considered. Scores have been
computed for vector wind at 200, 500 and 850\,hPa. Results are listed
in Tables~\ref{tab:deltas8to100} and \ref{tab:deltas16to100} under
configurations uv200L2, uv500L2 and uv850L2.  The score adjustment in
the fair logarithmic score works well overall for vector wind. As for
all previously shown results the adjustment is very precise $(|R| \le
0.03)$ at Day~2 and a bit less accurate 
$(|R| \le 0.08)$ at Day~5
(Tab.~\ref{tab:deltas8to100}). Furthermore, there is also an indication that the
adjustment at Day~5 works slightly better at 200 and 500\,hPa than at 850\,hPa.

\subsection{Profiles of vector wind}
\label{subs4.4}

\begin{figure*}[t]
\makebox[0.5\linewidth]{(a)}\makebox[0.5\linewidth]{(b)}
\includegraphics[width=0.49\linewidth]{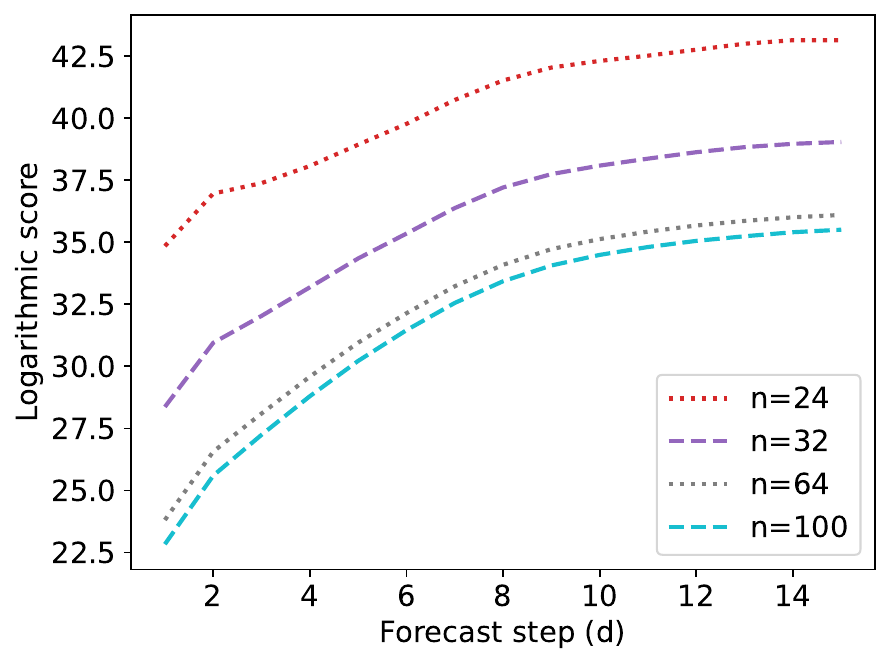}%
\includegraphics[width=0.49\linewidth]{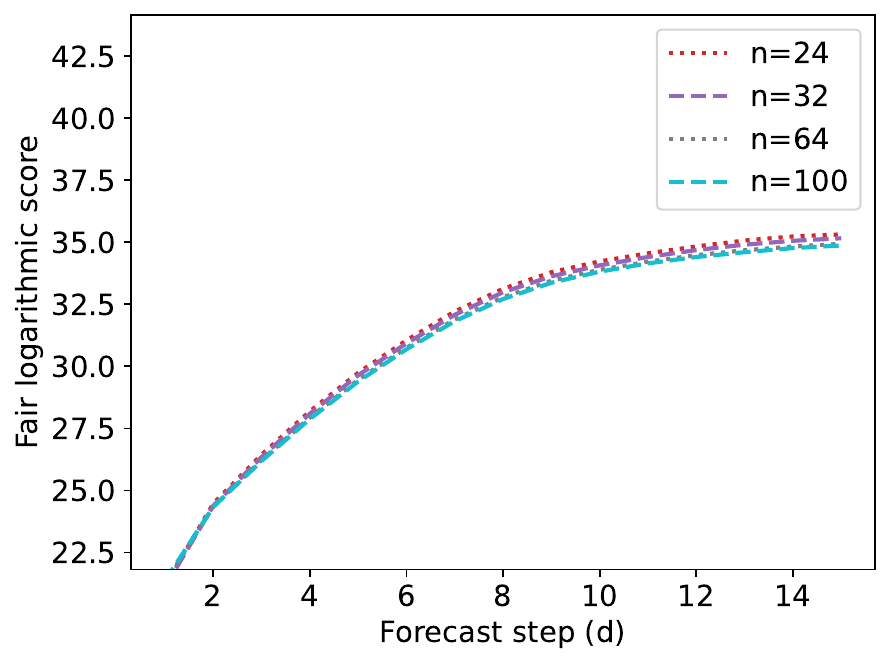}
  
  \caption{Ensemble size dependence of (a) logarithmic score and (b) fair logarithmic score for the 12-dim predictand of horizontal vector wind on the levels of 200, 300, 500, 700, 850, 925\,hPa  (configuration  uv200to925L12). The ensemble size ranges from $n=24$ to $n=100$.}
\label{fig:uv200to925L12}
\end{figure*}

Lastly, we consider forecasts consisting of horizontal vector wind at multiple levels. Figure~\ref{fig:uv200to925L12} shows the logarithmic scores of the 12-dim predictand of vector wind at 6 pressure levels between 200 and 925\,hPa. This is the highest-dimensional example considered in this study and it exhibits
the largest sensitivity to ensemble size of all the 13 considered configurations. The 15-day 100-member ensemble forecast has a lower
logarithmic score than the 2-day 24-member ensemble forecast
(Fig.~\ref{fig:uv200to925L12}a); nevertheless, the score adjustment again reduces the
sensitivity to ensemble size to a large extent
(Fig.~\ref{fig:uv200to925L12}b). Consistent with all earlier results,
the adjustment is nearly perfect for early lead times up to Day~2 and
than degrades slightly with increasing lead time. Scores have been
computed also for a 16-member ensemble. The logarithmic scores reach
values of about 90 and have been excluded from the figure as this
would have required an excessive range for the y-axis. However, the
score adjustment still works well despite having to make large
corrections (cf.\ configuration uv200to925L12 in
Tab.~\ref{tab:deltas16to100}).

Scores have been computed as well for the 4-dim predictand of vector
wind at 200 and 850\,hPa (configuration uv200to850L4). Results for
this configuration are included in Tables~\ref{tab:deltas8to100} and
\ref{tab:deltas16to100}. The score adjustment works also well for this
lower-dimensional predictand and is not worse than the adjustment of
vector wind at single levels.

\subsection{Deviation from normality}
\label{subs4.5}
It is expected that ensemble forecasts generally sample distributions that do not follow a multivariate Gaussian distribution. In order to quantify the level of deviation from normality, the Henze-Zirkler test statistic introduced in Section~\ref{subs3.3} has been computed for all 100-member ensemble forecasts, and the corresponding Wald test statistic \eqref{wald} has been averaged over all forecast start dates and all grid points in the northern mid-latitudes. In addition, multivariate normal distributions have been sampled in order to obtain reference values for the null hypothesis.

\begin{figure*}
    \centering
    \makebox[0.5\linewidth]{(a)}\makebox[0.5\linewidth]{(b)} \includegraphics[totalheight=0.38\linewidth]{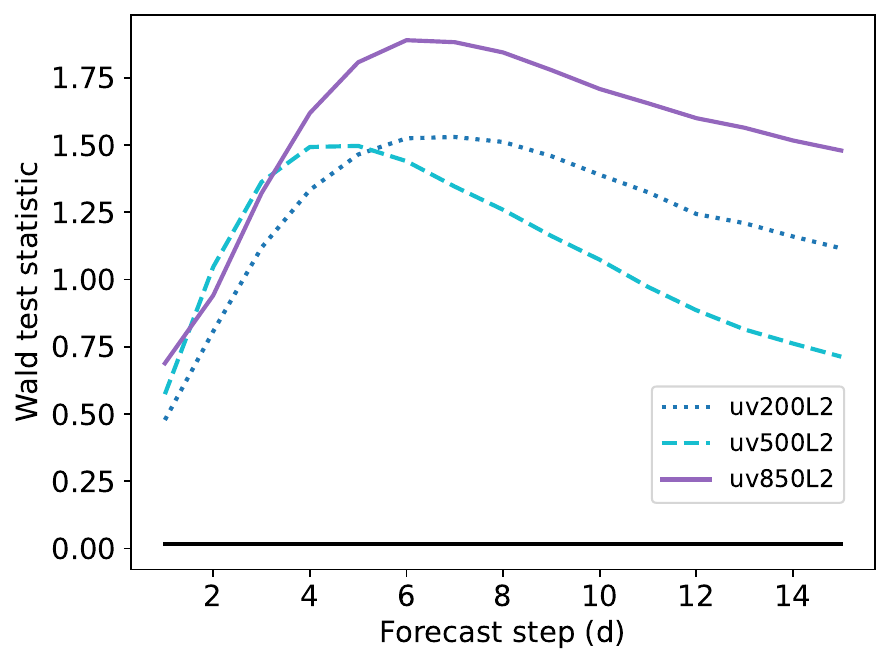} \includegraphics[totalheight=0.38\linewidth]{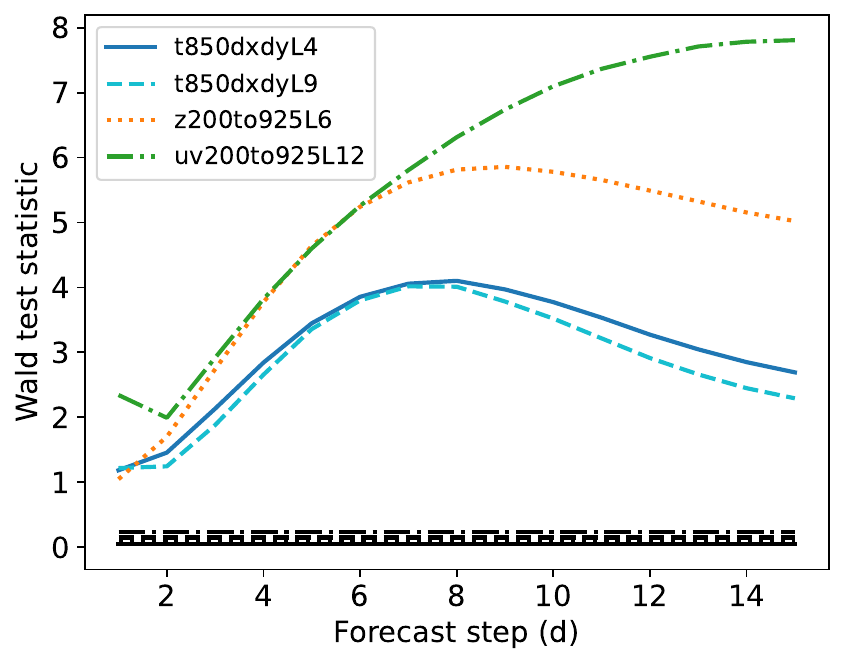}
    \caption{Wald test statistic based on the Henze-Zirkler test for multivariate normality.  The values are computed from 100-member ensemble forecasts and are averaged over start dates and the Northern mid-latitudes verification region. Panel (a) shows results for vector wind at single levels and panel (b) for the four configurations shown in Figures 2--5. Also shown are 
    numerical estimates of the mean Wald test statistic for the null hypothesis of a $p$-variate normal distribution (horizontal black lines with the same line style  as the NWP data).}
    \label{fig:wald}
\end{figure*}

The deviation from normality varies with the particular configuration of the predictands and the forecast lead time (Fig.~\ref{fig:wald}).  
The values of the Wald test statistic are smallest at early lead times and then rapidly grow. Several configurations of predictands have the largest deviations from normality in the medium-range between Day~4 and Day~8 and then become again somewhat closer to Gaussian distributions. An exception is the configuration with the vector wind profile uv200to925L12, which after Day~1 results in a monotone increasing Wald test statistic with a maximum  
at a lead time of 15\,d. The configurations of wind speed on the stencils also exhibit monotonously growing deviations from normality up to Day~15 (not shown).

The fair logarithmic scores (Figs.~\ref{fig:t850dxdyL4}--\ref{fig:uv200to925L12}, Tabs.~\ref{tab:deltas8to100}, \ref{tab:deltas16to100}) show the least dependence on ensemble size at early lead times, when deviations from normality are smallest, as one might expect. Vice versa, large deviations from normality, as quantified by the Wald test statistic, tend to coincide with slightly larger ensemble size dependence of the fair logarithmic score. For instance, configuration z200to925L6 exhibits a peak of the Wald test statistic at Day~8, and this coincides roughly with the largest ensemble size dependence of the fair logarithmic score (Fig.~\ref{fig:z200to925L6}b). Likewise the maximum Wald values for configuration uv200to925L12 at Day 15 coincide with the largest spread among ensemble sizes of the fair logarithmic scores (Fig.~\ref{fig:uv200to925L12}b).

\begin{table*}[t]
  \caption{Mean Wald test statistic values for $p$-variate uniform distributions. Values have been estimated numerically from a sample size of $10^5$.}
\centering

\medskip
  \begin{tabular}{c|cccccc}
  \hline
    $p$ & 2 & 3 & 4 & 6 & 9 & 12 \\ \hline
    $\langle\text{Wald}\rangle$ &
     3.225 & 3.450 & 3.423 & 3.039 & 2.435 & 2.053 \\
 \hline
  \end{tabular}
  \label{tab:wald-uniform}
\end{table*}

In order to gauge how large the deviations from normality hare in the different configurations, we have also computed mean Wald test statistic values for $p$-variate uniform distributions (Tab.~\ref{tab:wald-uniform}).  These are in the range from 2 to 3 depending on the dimension $p$. Therefore, we conclude that mean Wald values exceeding values of about 2 would suggest substantial deviations from normality. These occur in several configurations except for vector wind at single levels. Finally, note also that the condition \ $|Z|\geq 1.96$ \ indicates a significant deviation from normality  for individual forecasts at the 5\,\%  level since  the Wald test statistic \ $Z$ \ follows a standard Gaussian law under multivariate normality.

\section{Discussion}
\label{sec5}
Scores play a role when developing ensemble forecast systems. In order to increase the computational efficiency of the development process it is beneficial to test ensemble configurations with fewer members than in the operational ensemble. In this context, the ensemble size dependence matters even if one always compares ensembles forecasts with the same number of members.  \cite{siegert.ea.2019} emphasize this point for the logarithmic score in the univariate case. They show that the ensemble variance minimising the logarithmic score depends on ensemble size. For small ensembles an overdispersive ensemble appears superior to an ensemble with the correct amount of variance. To avoid problems due to extrapolating results obtained with small ensembles, fair scores are crucial.   They permit to correctly predict the score differences of ensemble configurations in the limit of large ensemble size from numerical experiments that have only a few members.

\begin{figure*}[!ht]
    \centering
\makebox[0.5\linewidth]{(a)\quad$n=6$}\makebox[0.5\linewidth]{(b)\quad $n=12$}
    \includegraphics[width=0.49\linewidth]{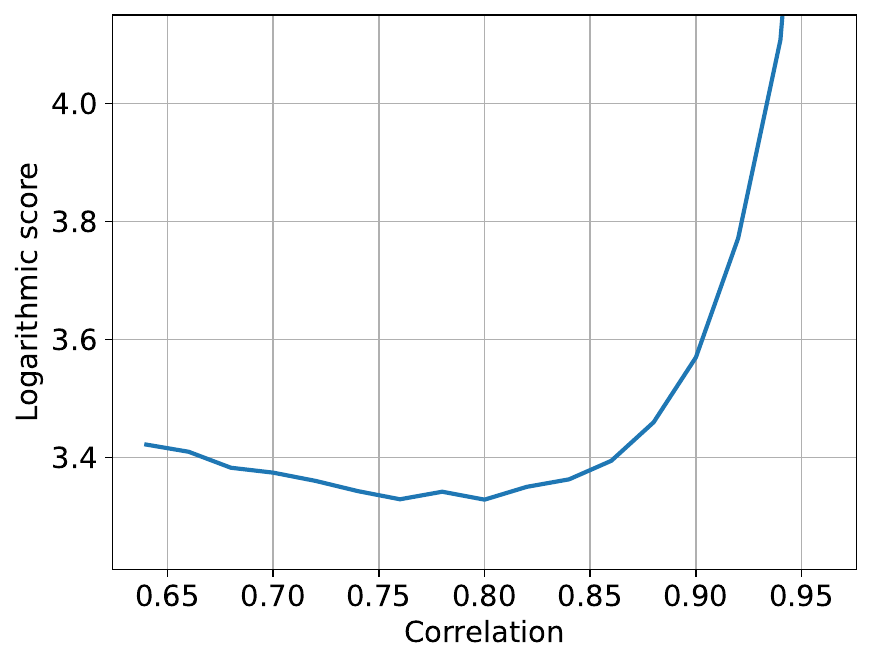}%
    \includegraphics[width=0.49\linewidth]{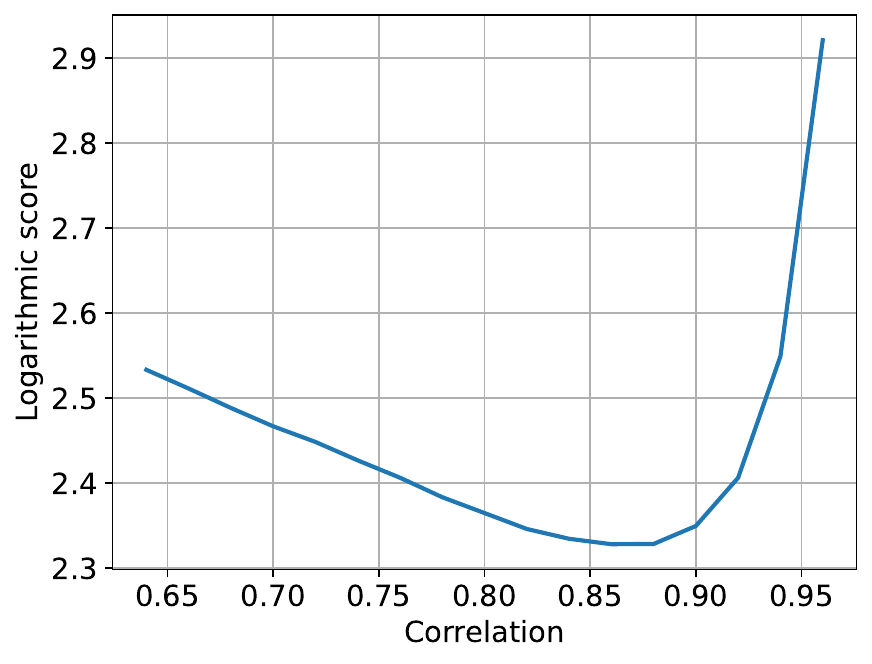}
\makebox[0.5\linewidth]{(c)\quad$n=100$}\makebox[0.5\linewidth]{(d)\quad $n=6$}
    \includegraphics[width=0.49\linewidth]{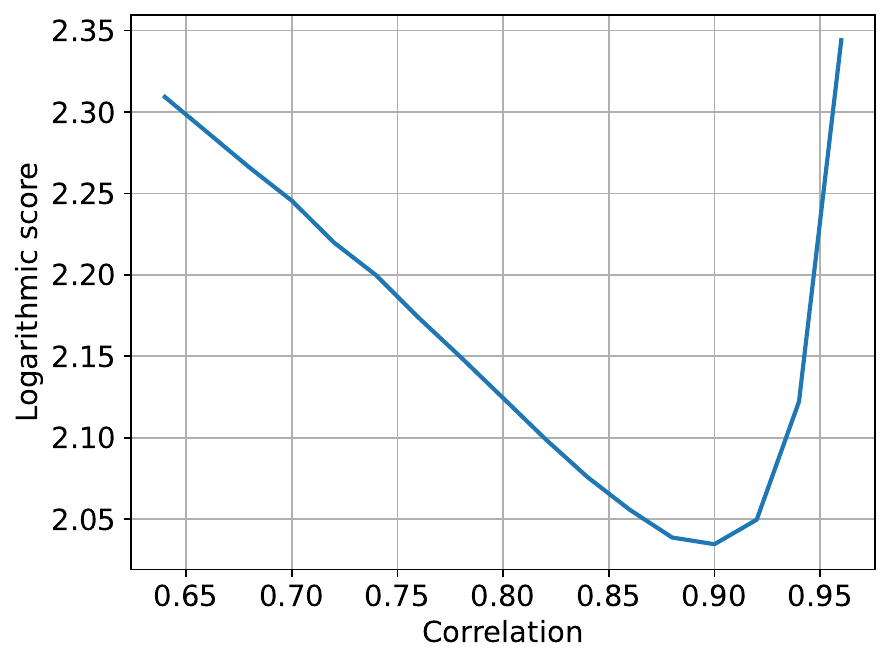}%
    \includegraphics[width=0.49\linewidth]{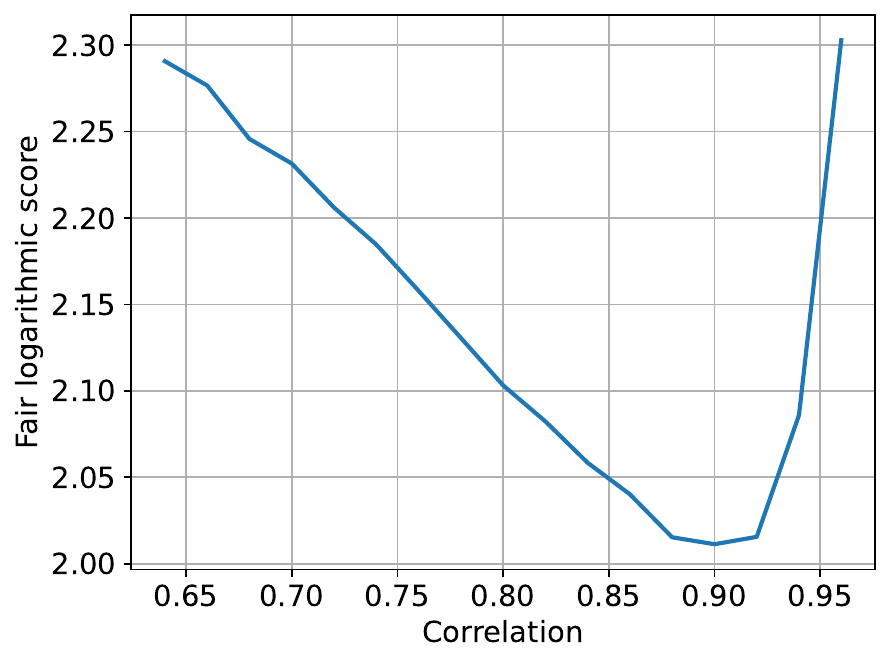}
    \caption{Logarithmic scores for $n$-member ensemble forecasts of a bivariate predictand. Panels (a)--(c) show the scores for ensemble sizes 6, 12 and 100, respectively.  Panel (d) shows the fair logarithmic score for a 6-member ensemble. The members have perfectly reliable marginal distributions with unit variance. The observed correlation of the components of the predictand is 0.9. The correlation of the components in the sampled predictive distribution is varied between 0.64 and 0.96. Scores are estimated numerically from a sample of  $10^6$ ensemble forecasts for each correlation value.}
    \label{fig:miscali}
\end{figure*}

When comparing ensemble skill with the logarithmic score for multivariate predictands this is equally important. To illustrate this, let us consider an ensemble prediction of a 2-dimensional vector. We assume that the sampled marginal distributions of the vector components are perfectly reliable while the correlation between the components may have an error. Let \ $\varrho_o$ \ and \ $\varrho_f$ \ denote the correlations between the components for the observation \ $\boldsymbol{y}$ \ and the forecasts \ $\boldsymbol{x}$, \ respectively. We assume that observations and forecasts sample the bivariate normal distributions
\begin{equation*}
  \boldsymbol{x} \sim {\mathcal N}_2\left( \boldsymbol{\mu},\begin{pmatrix} 1 &\varrho_f \\ \varrho_f &1\end{pmatrix}\right)\qquad\text{and} \qquad
    \boldsymbol{y} \sim {\mathcal N}_2\left( \boldsymbol{\mu},\begin{pmatrix} 1 &\varrho_o \\ \varrho_o &1\end{pmatrix}\right).
\end{equation*}
Figure~\ref{fig:miscali}a--c shows the logarithmic score for different predictive correlations $\varrho_f$ and three ensemble sizes when \ $\varrho_o=0.9$. \ The minimum of the logarithmic score  is at \ $\varrho_f=\varrho_o$ \ only for the 100-member ensemble. The 6 and 12-member ensembles have the score minimum at weaker correlations. In contrast, the fair logarithmic score computed from just 6 members has the score minimum at the correct correlation (Fig.~\ref{fig:miscali}d).

\citet{wilks.2017} examined metrics for assessing multivariate calibration and found that in many cases of miscalibration the Gaussian Box ordinate transform (BOT) proposed by \citet{gneiting.ea.2008} was the most sensitive metric. However, \citet{wilks.2017} cautioned that a large ensemble size $n\gg p$ is required for the practical usability of the BOT. 
We expect that the logarithmic score will exhibit a similar sensitivity to miscalibration as the BOT as the BOT simply assesses the distribution of the reliability term of the logarithmic score (see Section \ref{subs2.1}). As shown here, the demand on large ensemble sizes identified by \citet{wilks.2017} can be dropped when the fair logarithmic score is used as metric. Thus, the fair logarithmic score is expected to be a practical metric to monitor multivariate calibration together with sharpness.

This study has highlighted how the logarithmic score scales with the dimension \ $p$ \ of the predictand and ensemble size \ $n$ \ (cf.\ Fig.~\ref{fig:delta-logs}). With increasing dimension of the predictand, ensemble size has to increase quadratically to achieve the same score improvement in the logarithmic score. It may be challenging to drastically increase ensemble size beyond 100 for traditional NWP ensembles. Multivariate postprocessing techniques \citep{schefzik.moeller.2018,lerch.ea.2020} could help to better estimate dependency structures for higher dimensional predictands.
It is an interesting research question whether these methods might help to boost skill as measured with the logarithmic score without the need to increase ensemble size quadratically as the dimension of the predictand increases.

\section{Conclusions}
\label{sec6}
This work has been concerned with the logarithmic score for vector predictands and its dependence on ensemble size. An adjustment to the logarithmic score has been derived under the assumption that the ensemble forecasts can be considered independent realisations from a multivariate normal distribution and the score is computed from the normal distribution with mean and covariance determined by the ensemble forecasts. This generalises earlier work by \citet{siegert.ea.2019} for scalar predictands. The adjustment, which involves a multiplicative correction as well as additive ones, can be used to estimate the logarithmic score of an $N$-member ensemble from the logarithmic score of an $n$-member ensemble. When the limit \ $N\rightarrow\infty$ \ is considered, one obtains the fair logarithmic score for $p$-dimensional vector predictands under the assumption of normality.

In numerical weather prediction (NWP), ensemble forecasts are expected to exhibit considerable deviations from normality. Therefore, the derived adjustment of the logarithmic score can only approximate the actual ensemble size dependence. In order to judge its usefulness in such situations, a range of different multivariate predictands with dimensions from 2 to 12 have been considered. The data comes from the extended-range forecasts of ECMWF. These forecasts are initialised daily and have 100 perturbed members which are to first order exchangeable. Forecasts initialised in boreal autumn 2023 have been verified and results are focussed on the mid-latitudes of the northern hemisphere. The ensemble size is varied between 8 and 100 by selecting subsets of members. Forecasts of instantenous values up to Day 15 are considered here.

Results demonstrate how sensitive the logarithmic score is to ensemble size, in particular, for the higher dimensional predictands. The ensemble size adjustment works extremely well at early lead times up to Day 2 and is still considered good enough for practical use at longer lead times. In order to quantify how large the deviations from multivariate normality are in the NWP ensemble forecasts, a statistic proposed by \citet{henze.zirkler.1990} has been computed. This diagnostic shows that deviations from normality quickly increase with lead time peaking somewhere between Day 4 and Day 15 depending on the considered predictand. The largest deviations from normality were observed for vector wind profiles $(p=12)$, whereas the smallest ones appeared for vector wind at a single pressure level $(p=2)$ at early lead times. Although substantial deviations from normality are diagnosed, the ensemble size adjustment still approximates the true ensemble size dependence well.

Here the focus has been on the application of the ensemble size adjustment of the logarithmic score to the verification of medium-range weather forecasts. However, the methodology is generic and could be exploited in many other fields. Given its robustness in situations with some deviation from normality, it is expected to be of wider interest. It may become a useful tool in model development, in particular, for computationally demanding models where large ensemble are not desirable. In such situations, using the fair logarithmic score together with moderate ensemble sizes promises a more efficient development process due to savings in compute resources and faster  turn-around-times. 

\section*{Acknowledgements}
S\'andor Baran was supported by the Hungarian National Research, Development and Innovation Office under Grant K142849. He is also grateful to the ECMWF for supporting his research stay in Reading, United Kingdom. All figures have been produced with matplotlib \citep{Hunter.2007}.




\bibliographystyle{rss}

\bibliography{refsarXiv}

\end{document}